\documentclass[prl,preprint,showpacs,preprintnumbers,amsmath,amssymb,superscriptaddress]{revtex4}
\usepackage{graphicx}

\begin{document}
\title{Isotope shift in the dielectronic recombination of three-electron $^{\textrm{A}}$Nd$^{57+}$}
\author{C.~Brandau}
\email[e-mail:~]{c.brandau@gsi.de}
\author{C.~Kozhuharov}
\affiliation{Gesellschaft f\"{u}r Schwerionenforschung (GSI),
64291 Darmstadt, Germany}
\author{Z. Harman} \affiliation{Max-Planck Institut f\"ur Kernphysik, 69117 Heidelberg, Germany}
\author{A.~M\"uller}
\author{S.~Schippers} \affiliation{Institut f\"ur Atom- und Molek\"ulphysik,
Justus-Liebig-Universit\"at, 35392 Giessen, Germany}
\author{Y.S.~Kozhedub} \affiliation{Department of Physics, St.~Petersburg State University, 198504 St.~Petersburg, Russia}
\author{D.~Bernhardt}
\author{S.~B\"ohm}
\author{J.~Jacobi}
\author{E.W.~Schmidt}
\affiliation{Institut f\"ur Atom- und Molek\"ulphysik,
Justus-Liebig-Universit\"at, 35392 Giessen, Germany}
\author{P.H.~Mokler} \affiliation{Max-Planck Institut f\"ur Kernphysik, 69117 Heidelberg, Germany} \affiliation{Institut f\"ur Atom- und Molek\"ulphysik,
Justus-Liebig-Universit\"at, 35392 Giessen, Germany}
\author{F.~Bosch}
\author{H.-J. Kluge}
\author{Th.~St\"ohlker}
\author{K.~Beckert}
\author{P.~Beller}
\author{F.~Nolden}
\author{M.~Steck}
\author{A.~Gumberidze}
\author{R.~Reuschl}
\author{U.~Spillmann}
\affiliation{Gesellschaft f\"{u}r Schwerionenforschung (GSI),
64291 Darmstadt, Germany}
\author{F.J.~Currell} \affiliation{Physics Department, Queen's University, Belfast BT7 1NN, UK}
\author{I.I.~Tupitsyn}
\author{V.M.~Shabaev} \affiliation{Department of Physics, St.~Petersburg State University, 198504 St.~Petersburg, Russia}
\author{U.D.~Jentschura}
\author{C.H.~Keitel}
\author{A.~Wolf} \affiliation{Max-Planck Institut f\"ur Kernphysik, 69117 Heidelberg, Germany}
\author{Z.~Stachura}
\affiliation{Instytut Fizyki J\c{a}drowej, 31-342 Krak\'{o}w,
Poland}
\date{\today}
\begin{abstract}
Isotope shifts in dielectronic recombination spectra were studied for
Li-like $^{\textrm{A}}$Nd$^{57+}$ ions with A=142 and A=150.
From the displacement of resonance positions energy shifts
$\delta E^{142,150}(2s-2p_{1/2})=$ 40.2(3)(6) meV ((stat)(sys)) and $\delta E^{142,150}(2s-2p_{3/2}) = $ 42.3(12)(20) meV
of $2s-2p_{j}$ transitions were deduced. An evaluation of these values within a full QED treatment yields a change in the mean-square charge radius of $^{142,150}\delta \langle r^2 \rangle =$ -1.36(1)(3) fm$^{2}$. The approach
is conceptually new and combines the advantage of a simple atomic structure
with high sensitivity to nuclear size.
\end{abstract}

\maketitle
\begin{figure}
\begin{center}
\includegraphics[width=\columnwidth]{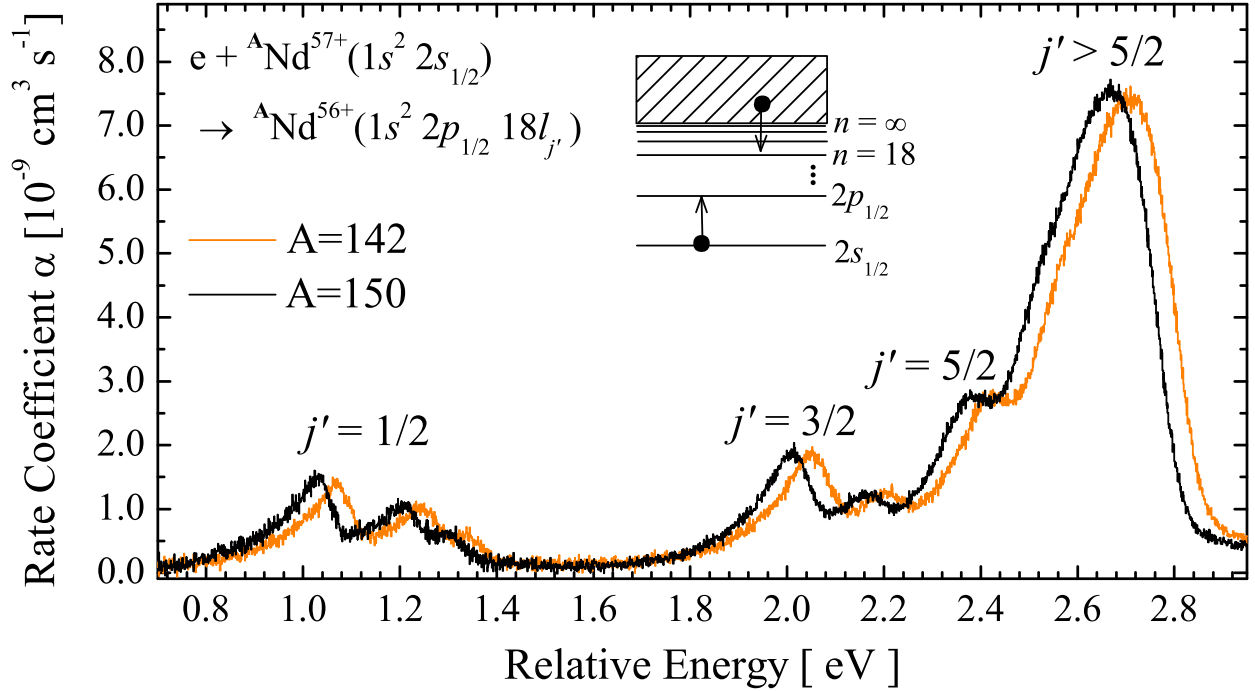}
\end{center}
\caption{Dielectronic recombination of the Li-like neodymium isotopes $^{142}$Nd$^{57+}$ (orange line) and $^{150}$Nd$^{57+}$ (black line)
in the energy range of the $1s^{2}\,2p_{1/2}\,18l_{j^{\prime}}$ resonance groups. The labels indicate the individual fine structure components $j^{\prime}$ of the $n=18$ Rydberg electron.}
 \label{fig:isoshift_n18}
\end{figure}
\begin{figure}
\includegraphics[width=\columnwidth]{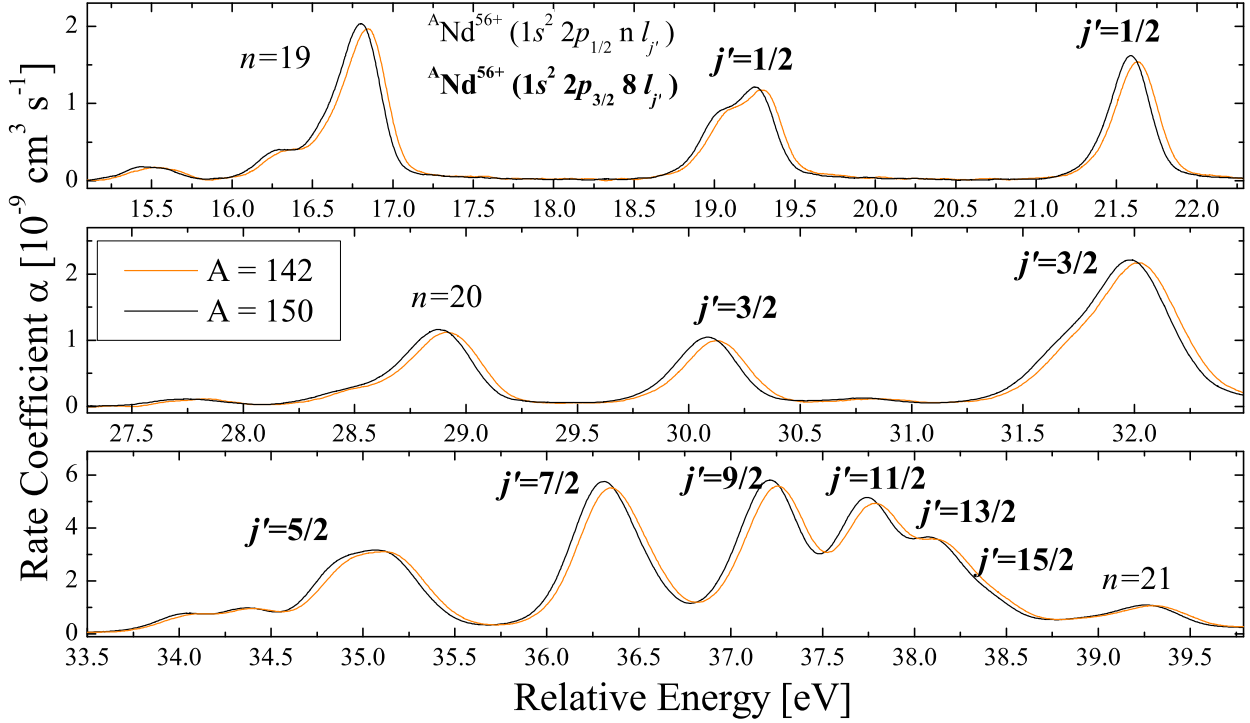}
\caption{\label{fig:isoshift_higher_energies}
Dielectronic recombination of A=142 (orange) and A=150 (black) Li-like $^{\textrm{A}}{\textrm{Nd}}^{57+}$. The labels $n$ indicate the principal quantum number of the Rydberg electron of the $1s^{2} \, 2p_{1/2} \, n l_{j^{\prime}}$ configurations and the bold labels $j^{\prime}$ denote the fine structure components of the $1s^{2} \, 2p_{3/2} \, 8 l_{j^{\prime}}$ resonances.}
\end{figure}
\begin{figure}
\begin{center}
\includegraphics[width=\columnwidth]{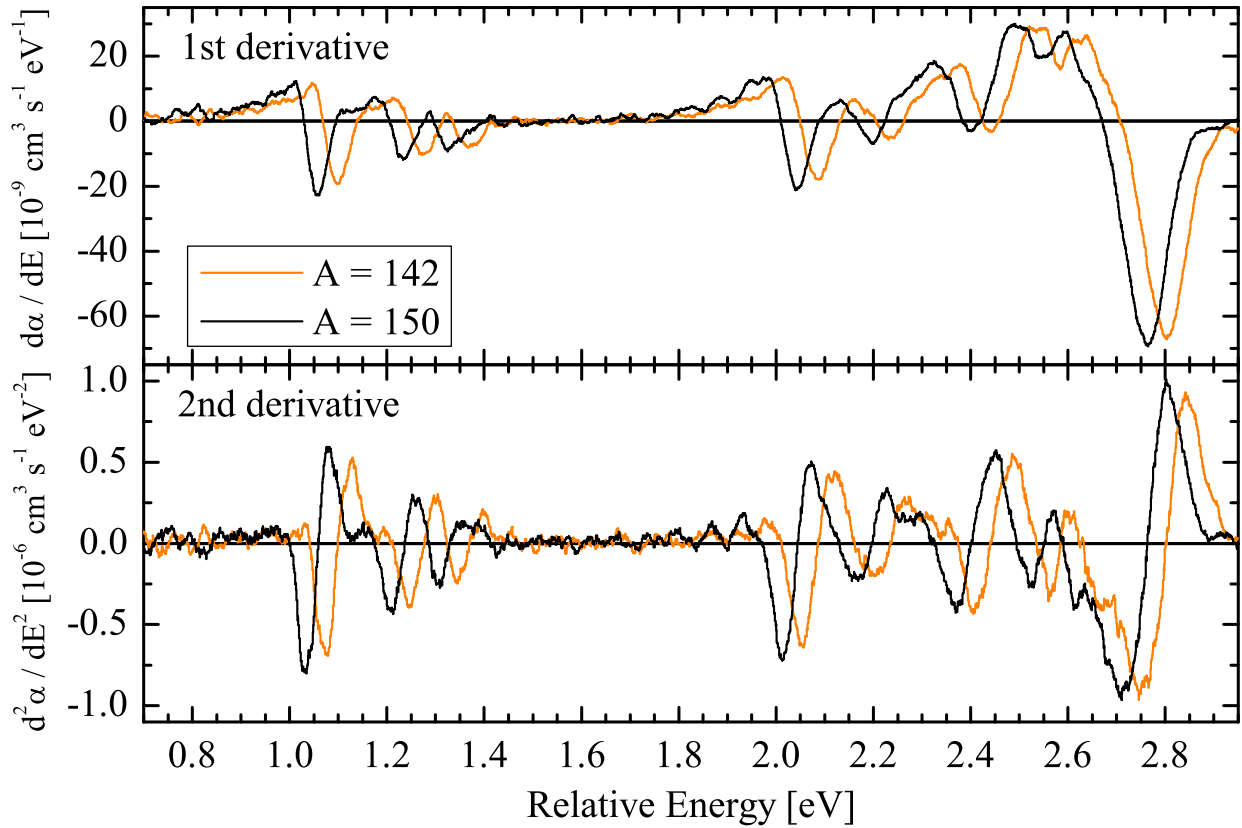}
\end{center}
\caption{First (upper panel) and second (lower panel) derivatives of the DR resonance spectra of the $^\textrm{A}$Nd$^{56+}$$1s^{2}\,2p_{1/2}\,18l_{j^{\prime}}$ group. The energy range is the same as in Fig.~\ref{fig:isoshift_n18}.
}
 \label{fig:isoshift_derivatives}
\end{figure}

The extent and shape of the proton distribution are basic properties of the atomic nucleus and reflect the interplay of forces that act between the nuclear constituents, protons and neutrons. The gross properties of  nuclear charge radii across the chart of nuclides---and also the fine details within isotope or isotone chains---provide primary observables for testing nuclear models \cite{Cook2006, Jain1990, Bender2003}. Charge radii, nuclear shape and hyperfine structure studies with classical methods such as electron scattering and x-ray spectroscopy of electronic and muonic atoms are essentially restricted to stable nuclei. Thus those isotopes far off the valley of stability cannot be addressed where the predictive power of nuclear models is poor, and interesting phenomena, like neutron or proton skins and shape effects occur.
This limitation is overcome in optical isotope shift (OIS) measurements which are well suited to study long isotopic chains including exotic nuclides \cite{Otten1989}. OIS is highly sensitive and possesses an enormous experimental precision. Since measurements of OIS are usually limited to low charge states, the interpretation of the data is, however, hampered by atomic many-body effects. The difficulty of a reliable theoretical description of
these complex electronic configurations results in large uncertainties due to specific mass shift contributions and in the calculation of the change of the electronic density at the site of the nucleus.\\
Detailed knowledge about nuclear properties and their influence on electron binding is essential for many applications in modern atomic physics research, in quantum chemistry \cite{Andrae2000} and even in the determination of chemical abundances of stellar objects \cite{Huehnermann1993}. This is particularly true for high-precision tests of fundamental interactions like atomic parity non-conservation \cite{Maul1996, Fortson1990, Dzuba2005} or quantum electrodynamics (QED) in strong electromagnetic fields \cite{Brandau2003, Beiersdorfer2005, Gumberidze2005}. Nuclear effects play a critical role in such tests and limit their explanatory power. The constraints are increasingly more severe for heavier elements since the scaling of nuclear effects with the nuclear charge Z exceeds the one of the sought-after contributions \cite{Mohr1998, Beier1998a}.\\
Radii compilations such as ~\cite{Otten1989, Fricke1995, Angeli2004} are widely used for reference. A closer look into the publications that form the basis of the compilations very often reveals inconsistencies.
For instance, for the present case of neodymium, more than 20 publications from all four classical techniques can be found (for references see \cite{Otten1989, Fricke1995, Angeli2004}). In \cite{Madsen1971} a decreasing mean-square charge radius for a change from isotope $^{142}$Nd to $^{150}$Nd is reported, whereas in \cite{Maas1974} a radius increase and a value of only $^{142,150}\delta \langle r^2 \rangle =$ -0.22 fm$^{2}$ are reported. While these values are considered as outliers the bulk of the data still covers a range from about $^{142,150}\delta \langle r^2 \rangle =$ -1.20 fm$^{2}$ to -1.36 fm$^{2}$.\\
In view of the wide range of applications and the scatter of the experimental data alternative methods for a reliable determination of charge radii are highly desirable. In this Letter we present a conceptually different experimental access to charge radii changes that is particularly well suited for the investigation of heavy stable and also unstable nuclides. The method is based on the storage ring measurement of isotope shifts (IS) $\delta E^{\textrm{A},\textrm{A}^{\prime}}$ in the spectrum of resonant electron-ion recombination of heavy few-electron ions. The technique is demonstrated for the case of the two stable even-even isotopes $^{142}$Nd$^{57+}$ and $^{150}$Nd$^{57+}$ of Li-like neodymium. Storage rings provide very clean experimental conditions since isotopically pure, quasi-monoenergetic (``cooled") beams in a single charge state are available. The experiments were performed at the electron cooler of the heavy ion storage ring ESR \cite{Franzke1987} at GSI in Darmstadt, Germany.
The benefit of using few-electron ions for IS studies was previously shown employing classical spectroscopy at electron-beam ion traps (e.g., \cite{Elliott1996, SoriaOrts2006}). Due to the simplicity of the atomic configurations, the interpretation of the data is clear and without ambiguity. For the $2s \to 2p$ transitions of Li-like ions the electronic part can be treated theoretically with high accuracy. Many-body and mass effects are small and can be reliably accounted for.\\
Resonant electron-ion recombination---also termed dielectronic recombination (DR)---can be viewed as a two-step process. The first step of DR is called dielectronic capture (DC) and is time-reverse to autoionization. Free electrons that posses matching kinetic energy can recombine resonantly and excite a bound electron. If---in a second step---DC is succeeded by emission of photons,
DR is complete. For the $2s \to 2p_{j}$ ($j = 1/2, 3/2$)
excitations of $^{\textrm{A}}$Nd$^{57+}$ DR can be described by
\begin{eqnarray}
  e^- + \, ^{\textrm{A}}\textrm{Nd}^{57+}(1s^{2} \, 2s_{1/2}) & \to & ^{\rm A}{\textrm{Nd}}^{56+}(1s^{2} \, 2p_{j} \, n l_{j^{\prime}})^{**}
  \nonumber \\
   & \to & ^{\textrm{A}}{\textrm{Nd}}^{56+}+ \mbox{photons}.
  \label{eq:DR}
\end{eqnarray}
In Nd$^{57+}$, DC can proceed via Rydberg states with principal quantum numbers $n\geq 18$ for the $2s \to 2p_{1/2}$ core excitations and $n \geq 8$ for $2s \to 2p_{3/2}$, respectively (compare Figs.~\ref{fig:isoshift_n18} and \ref{fig:isoshift_higher_energies}). With increasing electron-ion collision energy, series of Rydberg resonances up to the series limits $n \to \infty$ are formed \cite{Brandau2002, Brandau2003}. For different isotopes, the whole resonance series exhibits the same energy shift, i.e., the one of the corresponding core excitation. The captured loosely bound Rydberg electron itself is almost unaffected by the small variations of the nuclear potential and its contribution to the isotope shift can be safely neglected.\\
In recent DR experiments at the storage rings CRYRING in Stockholm and TSR in Heidelberg, the high resolution that can be achieved for exceptionally low-energy resonances was exploited to explore magnetic hyperfine effects in Cu-like $^{207}$Pb$^{53+}$ \cite{Schuch2005} and in Li-like $^{45}$Sc$^{18+}$ \cite{Wolf2007}.
Low-lying resonance are well known for selected ions of intermediate charge state but are unlikely for heavy few-electron ions.
The Li-like ions under study in this Letter are ideal candidates for precision DR-IS experiments and provide an explicit and straightforward access to nuclear parameters.
The $2s$ electronic wave function and the nucleus posses large mutual overlap leading to strong nuclear size contributions. In any element, DR resonances with excitations  $2s \to 2p_{1/2}$ and in some cases additional resonances belonging to $2s - 2p_{3/2}$ transitions can be found in the favorable energy range below a few ten eV.\\
Detailed descriptions of the recombination set-up at the electron cooler of the ESR are given elsewhere \cite{Hoffknecht2001, Brandau2002, Brandau2003}. Here, we concentrate on the issues specific for the IS experiment with Li-like $^\textrm{A}$Nd$^{57+}$: Momentum selected $^\textrm{A}$Nd$^{57+}$ ions with A=142 or A=150 were injected into the storage ring ESR and stored at an energy of 56.3 MeV/u. In order to minimize lateral beam size and energy spread electron cooling was applied. For both isotopes an identical space-charge corrected cooling voltage of $U_{C}=30899$~V was chosen, equivalent to an ion velocity $\beta_{i}=0.3328$  in units of the speed of light. In addition to its normal function as a beam cooling device, we operated the electron cooler as a target of free electrons. After a series of recombination measurements with the reference isotope A=142, the operation parameters of the accelerator and the storage ring were scaled with the known mass ratio to the new isotope A=150. Special care was taken to ensure the same alignment of the ion beam with respect to the electron beam. Non-zero collision energies were introduced by applying a sequence of swiftly modulated voltage steps $U_{d}$ to a cylindrical drift tube located in the overlap region of the two beams. The relative collision energy $E_{cm}$ in the center-of-mass system can then be easily inferred from the Lorentz factors $\gamma_{i} - 1 =  e U_{C} / (m_{0,e}c^{2}) $ of the ion and $\gamma_{e} - 1 =  e ( U_{C} + U_{d} )/ (m_{0,e}c^{2}) $ of the electrons. Here, $m_{0,e}$ is the electron mass.
With  $-2.5$ kV $ \leq U_{d} \leq +2.5$ kV used in the present experiment, collision energies of $0 < E_{cm} \lesssim 50$ eV were covered. Recombined ions were separated from the primary beam in the next ESR bending magnet and were registered by a multi-wire proportional chamber. A typical voltage sequence contained measurements at $\sim$4000 different voltages of 33 ms  duration each. Intermediate cooling steps of the same duration guaranteed a constant ion energy and beam quality.
A constant electron current $I_{el} = 100$ mA was applied, corresponding
to a density $n_{e} = 3.1 \cdot 10^6$ cm$^{-3}$ at cooling. Comparable numbers of ions of typically about $N_{i}=2 \cdot 10^7$ for both isotopes were available in the ring. Recombination rate coefficients $\alpha$ were derived by normalizing the count rate using the primary beam intensities \cite{Brandau2002}.\\
Isotope shift data were taken for three different energy ranges,
$0 - 3.8$ eV, $12 - 24$ eV and $24 - 42$ eV with nominal step widths of 1 meV, 3 meV and  4 meV, respectively. With these settings, resonance groups $^{\textrm{A}}{\textrm{Nd}}^{56+}(1s^{2} \, 2p_{1/2} \, n l_{j^{\prime}})^{**}$ with $n = 18 \ldots 21$  and $^{\textrm{A}}{\textrm{Nd}}^{56+}(1s^{2} \, 2p_{3/2} \, 8 l_{j^{\prime}})^{**}$ were covered. The shift of resonance positions for the two isotopes is immediately evident from the spectra in Figs.~\ref{fig:isoshift_n18} and \ref{fig:isoshift_higher_energies} and is clearly visible over the full energy range. For the extraction of the IS values, maxima, minima and inflection points of the resonance spectra were used. By evaluating the shift of these characteristic points, errors due to normalization and background subtraction are minimized, and an analysis is achieved that is independent of supplementary input on the DR process and independent of the experimental response function.
First and second derivative spectra (Fig.~\ref{fig:isoshift_derivatives}) as well as smoothed versions of the original spectrum
were obtained from local fits of third-order polynomials for every point $i$ of the spectrum taking into account $m$ adjacent values on both sides of $i$. Full minimizations are performed accounting for a non-equidistant energy axis and for the error bars in both coordinates.
The procdure is similar to the Savitzky and Golay smoothing algorithm \cite{Savitzky1964} that retains fine details in the spectrum much better than averaging or re-binning \cite{Savitzky1964}.
For the analysis $m$=35 was used and no significant bias for $m$=25 or $m$=45 was found. For the IS determination only those roots of the derivatives were included that are significantly above the noise level. For the lowest energy range 7 independent IS data sets were obtained, 3 for the middle one and 1 for the  energy range 24 - 42 eV yielding in total 154 characteristic values for the $2s - 2p_{1/2}$ IS and 45 for the $2s - 2p_{3/2}$ IS. The average ISs are $\delta E^{142,150}(2s-2p_{1/2})=$ 40.2(3)(6) meV and a slightly higher value of $\delta E^{142,150}(2s-2p_{3/2}) = $ 42.3(12)(20) meV. Here, first and second parentheses denote statistical and systematic errors. Although statistically not significant, the experimentally observed small difference in the ISs is expected, since in the relativistic case the $p_{1/2}$-electron has a finite overlap with the nucleus. The main sources of systematic errors stem from ``slope" uncertainties, i.e., due to still remaining imperfections in background subtraction and normalization and from the finite digitalization of the energy steps. As a consequence of the merged-beams geometry, only minor contributions of below 0.2 meV for the $2p_{1/2}$ value and of $\sim 0.6$ meV for the $2p_{3/2}$ IS arise from a potential beam misalignment or voltage calibration uncertainties of $U_C$ and $U_{d}$.\\
The IS values were evaluated within two different fully relativistic atomic structure methods, namely, the multiconfiguration Dirac-Fock (MCDF) and
the configuration interaction Dirac-Fock-Sturmian methods (see \cite{SoriaOrts2006} and references therein). The nucleus was modeled with a two-parameter Fermi distribution $\rho(r) = N (1+\exp[(r-c)/a])^{-1}$, where $c$ is the half-density radius, $N$ is a normalization constant, and $t$= $4 a \ln 3$ the skin thickness \cite{Fricke1995}. IS measurements are basically insensitive to details of the charge distribution as long as the same root-mean-square (rms) radius is reproduced (e.g.~\cite{Otten1989, Fricke1995, Shabaev1993}). We employed $t=$ 2.3 fm for both isotopes in accordance to the established value used for heavy ions \cite{Shabaev1993, Fricke1995, Elliott1996, Mohr1998, Beier1998a}. The semi-magic isotope $^{142}$Nd (82 neutrons) served as a reference using the rms radius of $\langle r^2 \rangle^{1/2} = 4.9118$ fm from the most recent compilation of Angeli \cite{Angeli2004}. The radii values of Angeli were obtained from a combined analysis of comprehensive experimental input data from the four classical methods. The neutron-rich isotope $^{150}$Nd (N=90) is strongly deformed yielding a pronounced increase of the charge radius beyond the liquid drop model \cite{Otten1989}.\\
For the $2s - 2p$ transitions of $^{\textrm{A}}$Nd$^{57+}$ the total mass shift A=142$-$A=150 is as small as $\sim$4~\% of the IS and sums up to -1.63 meV for the $2s \to 2p_{1/2}$ excitation and to -1.80 meV for $2s \to 2p_{3/2}$, respectively. These values for the mass shift were obtained taking into account relativistic and QED recoil contributions \cite{Artemyev1995, SoriaOrts2006}: For the $2s-2p_{1/2}$ ($2s-2p_{3/2}$) transition the mass shift comprises of -2.44 meV (-2.53 meV) from averaging of the non-relativistic recoil operator with relativistic multiconfiguration wave functions, 1.14 meV (1.03 meV) from the relativistic recoil operator, and -0.33 meV (-0.30 meV) from the QED recoil effect. For the determination of $\delta \langle r^2 \rangle $, influences of nuclear size (NS) variations on Coulomb electron correlation, on the Breit interaction as well as on QED contributions were taken into account.
Similar to OIS \cite{Otten1989}, to a good approximation the IS depends linearly on $\delta \langle r^2 \rangle $, yielding a slope of -31.1 meV/fm$^2$ for the $2p_{1/2}$ and -32.3 meV/fm$^2$ for the $2p_{3/2}$ transition, respectively.
A further small correction that is often neglected \cite{Otten1989} arises from nuclear polarization (NP). It is sizeable for deformed nuclei with high nuclear transition probabilities B($\tau L$) such as $^{150}$Nd. Using the analytic formulae of \cite{Nefiodov1996} we estimate for the dominant contribution from the first excited 2$^+$ state at 130.21 keV (B(E2$\uparrow$)=2.760 e$^2$b$^2$, \cite{Raman2001}) a value of -0.3 meV. Finally, from the NP corrected value of $\delta E^{142,150}(2s-2p_{1/2})=$ 40.5(3)(6) meV, a change of $^{142,150}\delta \langle r^2 \rangle =$ -1.36(1)(3) fm$^{2}$ is derived. Consistently, we obtain $^{142,150}\delta \langle r^2 \rangle =$ -1.38(4)(7) fm$^{2}$ from the $2p_{3/2}$ resonances but with significantly larger error bars. The systematic errors include additional theoretical uncertainties of 0.01 fm$^2$, predominantly from NP. We have added the full
NP contribution to the error balance, since NP still lacks experimental verification. Our result of -1.36(1)(3) fm$^{2}$ is somewhat larger than the average experimental value of $^{142,150}\delta \langle r^2 \rangle =$ -1.291(6) fm$^{2}$ from the combined analysis in Angeli \cite{Angeli2004}. However, our values obtained with this alternative method are in good agreement with the upper end of the individual data that build the basis of this evaluation (see discussion on the first page) and support a larger increase of the radius from A=142 to A=150.\\
In this Letter a novel method for nuclear size studies was developed and
its general feasibility was demonstrated for heavy three-electron isotopes. These simple systems allow for a reliable analysis of the IS data within state-of-the-art structure calculations taking into account relativistic and QED contributions. ISs could be measured with $<$1 meV accuracy. For heavier ions a similar experimental precision can be retained. Due to the Z$^5$ to Z$^6$ scaling of nuclear size effects, for the heaviest ions ISs are of the order of 100 meV \cite{Brandau2003a} and hence the sensitivity of DR to the charge distribution is significantly enhanced. In contrast to OIS, isoelectronic studies of different elements are possible in order to disentangle atomic and nuclear contributions.
High experimental resolution and large resonant atomic cross sections enable future DR-IS experiments with unstable isotopes or long-lived nuclear isomers.  At GSI, such exotic highly charged ions can be synthesized in the fragment separator FRS \cite{Geissel1992} and subsequently investigated in the ESR. Thus, long chains with constant neutron number (isotone shifts) become accessible as well.
At the upcoming Facility for Antiproton and Ion Research (FAIR) \cite{Gutbrod2001} high production yields of radioactive ions and a dedicated electron target at the New Experimental Storage Ring (NESR), will furthermore increase accuracy, versatility and expand the scope of application.
\bibliographystyle{apsrev}

\clearpage

\end{document}